\documentclass[onecolumn]{emulateapj}
\usepackage{epsf}
\usepackage{bm}
\newcommand{\dpdp}[2]{\frac{\partial #1}{\partial #2}}

\begin{document}
\title{Generation of Magnetic Field on the Accretion Disk around a Proto-First-Star}
\author{Yuki Shiromoto, Hajime Susa}
\affil{Department of Physics, Konan University, Kobe 658-8501, Japan\altaffilmark{1}}
\author{Takashi Hosokawa}
\affil{Department of Physics and Research Center for the Early
Universe, The University of Tokyo, Tokyo 113-0033, Japan\altaffilmark{2}}
\email{susa@konan-u.ac.jp}
\begin{abstract}
The generation process of magnetic field around a
 proto-first-star is studied. 
 Utilizing the recent numerical result of proto-first-star
 formation based upon the radiation hydrodynamics simulations, we
 assess the magnetic field strength generated by the radiative force and the Biermann battery
 effect.

We find that magnetic field of $\sim 10^{-9}$ G is generated on the
 surface of the accretion disk around the proto-first-star.
 The field strength on the accretion disk is smaller by two orders of magnitude
 than the critical value, above which the gravitational fragmentation of the disk is suppressed.
 Thus, the generated seed magnetic field hardly affect the
 dynamics of on-site first star formation directly, unless efficient
 amplification process is taken into consideration.

 We also find that the generated magnetic field is continuously blown out from the disk on
 the outflows to the poles, that are driven by the thermal pressure of
 photoheated gas. 
 The strength  of the diffused magnetic field in low density regions
 is $\sim 10^{-14}-10^{-13}$G at $n_{\rm H}=10^{3}{\rm cm^{-3}}$
 which could play important roles on the next generation star formation,
 as well as the seeds of magnetic field exist in present-day universe.
\end{abstract}
\keywords{early universe---HI\hspace{-.1em}I regions ---radiative transfer --- magnetic fields}

\section{Introduction}
The energy density of magnetic field in local interstellar gas is
not negligible, which plays significant roles on the formation of stars in
our Galaxy. 
This magnetic field is generally regarded as a consequence of dynamo
amplification of initial seed magnetic field generated in the early
universe. Various generation mechanism of this seed field have been
proposed so far: magnetic field generation in the very early universe
due to the coupling of gravity and electromagnetic field\citep[e.g.][]{turner},
non-zero $\bm{\nabla}\times \bm{E}$ term due to the inhomogeneity of
radiation field at the recombination of the universe\citep[e.g.][]{ichiki}, the Biermann
battery effect\citep{biermann} at the shock front in forming galaxies\citep[e.g.][]{kulsrud,xu}/ at the
ionization fronts in the reionizing universe\citep{gnedin}. It has also been
proposed that the radiation force and the Biermann battery effect
in the neighbor of luminous sources can generate seed
field\citep{langer2,ando,doi}. Interestingly, all of these models suggest the seed field
strength of $10^{-20}-10^{-18}$G at IGM densities, that is well below the observational
constraint\citep[][and the references therein]{durrer_neronov}.

The generated seed magnatic fields could potentially affect the star
formation process in the early universe as seen in our Galaxy, e.g.,
through launching jets, transferring angular momentum, and suppressing
the gravitational fragmentation of disks. It has been pointed out that
the registivity of the primordial gas should be relatively low
throughout the collapse of a cloud \citep[e.g.][]{maki04,maki07,schleicher09}. As a result, magnetic fields do not dissipate
from the cloud at Jeans scale.
Thus, given
a same strength of seed magnetic field at the onset of the collapse,
magnetic field should has larger impact on star formation in the
primordial gas than that in the present-day molecular cloud in which
magnetic field dissipate within a certain range of
densities\citep[e.g.][]{nakano_umebayashi}.

\citet{machida08} have addressed the dynamical effects of magnetic field
on the star formation in primordial gas, assuming ideal MHD, which is
correct as far as we consider larger scales than the Jeans length. 
They found that the magnetic field do have
dynamical impact on the collapsing gas including the formation of jets
if $B\gtrsim10^{-9}$G at $10^{3}{\rm cm^{-3}}$. 
\citet{machida_doi13} addressed the later evolution of the system (the
gas accretion phase ) by resistive MHD calculations. 
They found that the fragmentation of the
disk is significantly suppressed by the magnetic field in case $B \gtrsim
10^{-10} (n_{\rm H}/10^3 {\rm cm}^{-3})^{-2/3} $G.
However if we convert this
field strength to the IGM density at $z=20$ assuming flux freezing, we
obtain $B\gtrsim 10^{-14}$G which is much larger than the seed field strength
predicted by various models as mentioned above. Hence, magnetic field
seems unlikely to affect the dynamics of the formation of primordial stars. 
To put it the other way around, if seed field of $B\gtrsim 10^{-14}$G is
generated by some mechanism, magnetic field is a vital part of star
formation in the early universe.
It has been suggested that small scale turbulence can
amplify the seed magnetic field with a dynamical time scale, and it
inversely cascade into larger scales to affect the dynamics of star
forming cloud at Jeans scale\citep{schleicher10,schober12}, although the
sufficient amplification to affect the dynamics of the gas has not been
shown starting from very week seed field of $\sim
10^{-18}-10^{-20}$G by ab initio numerical simulations so
far\citep{sur10,federrah11,turk12,latif13}. 
In any case the amplitude of seed magnetic field is an important
key parameter to understand the primordial star formation. 

In this paper, we discuss the seed magnetic field generation mechanism
regarding anisotropic radiation field and complex
density/temperature structures in the very neighbor of forming first
stars, which was originally discussed in our previous
studies at 100pc-10kpc scales \citep{ando,doi}.  
We focus on the magnetic field generation especially in the very
neighbor of proto-first-stars at 100-1000AU scale 
where accreting/outflowing gas create intricate structures.
Since the generated field strength depends on the radiation flux from
the source star, stronger magnetic field is expected.

Such complicated structures have already been calculated by \citet{hosokawa11}.
They investigated the gas accretion process onto  a proto-first-star
by two dimensional radiation hydrodynamics simulations, although the generation
process of magnetic field was not taken into account. 
In this paper, we evaluate the growth of magnetic field by
post-processing scheme utilizing their simulation as a background.
Then we discuss the impact of the magnetic field generated through
this process on the evolution of the proto-first-stars and the formation
of second generation stars.

This paper is organized as follows: we describe the fundamental processes
of seed field generation in section \ref{mag_generation}. In section \ref{model} we show
our numerical model, and the results are shown in
section \ref{results}. Section \ref{discussion} is devoted to
discussion and conclusion.

\section{Generation of magnetic field}
\label{mag_generation}
In this section, we briefly describe the basic equations of magnetic field
generation, more detailed description can be found in \citet{ando}
though. 
The growth of magnetic field is described by the following induction
equation with two source terms:

\begin{eqnarray}
\dpdp{\bm{B}}{t}=\bm{\nabla}\times\left(\bm{v}\times\bm{B}\right)-\frac{c}{e n_{\rm e}^2}\bm{\nabla}n_{\rm e}\times\bm{\nabla}p_{\rm e}-\frac{c}{e}\bm{\nabla}\times\bm{f}_{\rm rad}\label{eq:B_growth}
\end{eqnarray}

This equation is obtained by combining the equation of the force balance on a
single electron with the Maxwell's equations.
Here we omitted the Hall current term, which is proportional to
$\bm{j}\times \bm{B}$, since the this is the higher order term than the others, in
case we consider the generation of weak magnetic field. 

The first term in the R.H.S. denotes the advection of magnetic field
lines which stick to the gas. 
If the second and third term is neglected, frozen-in condition is
satisfied. Thus, the first term is not the source of
the magnetic field generation. The second term is the Biermann battery
term\citep{biermann}, which is non-zero in case the gradient of electron
density and pressure are not perfectly parallel with each other. The third term is the radiation
term, which is proportional to the rotation of the radiation force field
on a
single electron ($\bm{f}_{\rm rad}$). 

The radiation force $\bm{f}_{\rm rad}$ is composed of two micro
processes that could be potentially important for the momentum transfer
from photons to electrons. First one is the Thomson scattering.
We use formal solution of radiation transfer equation for
$\bm{f}_{\rm rad,T}$ as follows:
\begin{eqnarray}
\bm{f}_{\rm rad,T}  = \frac{\sigma_{\rm T}}{c}\int_0^{\nu_{L}}\bm{F}_{0\nu}d\nu + \frac{\sigma_{\rm T}}{c}\int_{\nu_{L}}^{\infty}\bm{F}_{0\nu}\exp\left[- \tau_{\nu_{L}}a\left(\nu\right)\right]  d\nu,\label{eq:momtr_T}
\end{eqnarray}
where $\bm{f}_{\rm rad,T}$ denotes the radiation force per single
electron due to Thomson scattering, 
$\sigma_{\rm T}$ is the cross section, 
$\bm{F}_{0\nu}$ is the incident energy flux density.
$\nu_{L}$ denotes the Lyman-limit frequency,
$\tau_{\nu_{L}}$ is the optical depth at the Lyman limit regarding the
photoionization, and $a(\nu)$ denotes the frequency dependence of
photoionization cross section, which is normalized at the Lyman limit,
i.e. $a(\nu_L)=1$ is satisfied.

The second process is the photoionization, whose contribution to $\bm{f}_{\rm rad}$ is given as
\begin{eqnarray}
\bm{f}_{\rm rad,I}  = \frac{1}{2}\frac{n_{\rm HI}}{c n_{\rm e}}
\int_{\nu_{L}}^{\infty} \sigma_{\nu_{L} } a\left(\nu\right)
\bm{F}_{0\nu}\exp\left[- \tau_{\nu_{L}} a\left(\nu\right)\right]  d\nu, \label{eq:momtr_ion}
\end{eqnarray}
where $\sigma_{\nu_L}$ denotes the photoionization cross section at the Lyman Limit. We remark that the factor $1/2$ in R.H.S. of
equation (\ref{eq:momtr_ion}) is due to the fact that the photon
momentum is equally delivered to protons and electrons. 
Then, we also can safely assume that only electrons are accelerated by
this momentum transfer process, since protons have much larger inertia than electrons.

\section{Model description}
\label{model}

\subsection{The underlying model of proto-first-star formation}
We take the underlying model of the proto-first-star formation from
\citet{hosokawa11}. The initial condition of the model is taken from
\citet{yoshida08}, where the formation of the embryo protostar is
simulated in a fully cosmological context. The subsequent evolution in
the mass accretion stage is followed with 2D radiation hydrodynamic
simulations coupled with the protostellar evolution calculations (see
Hosokawa et al. 2011 for full details). The stellar mass rapidly
increases in this stage, as the gas accretes onto the star through a
circumstellar disk. The stellar UV luminosity also sharply rises with
increasing the stellar mass, and a bipolar HII region emerges. The HII
region then begins to expand dynamically in the accreting envelope,
which blows away the gas in the envelope. The circumstellar disk is
exposed to the stellar ionizing radiation, and gradually loses its mass
owing to the photoevaporation. In the fiducial case, which is adopted
here,  a $43~M_\odot$ star forms as the mass accretion is finally shut
off by the stellar UV feedback effect.

\subsection{Magnetic field generation by postprocessing}
\begin{figure}[t]
\begin{center}
\includegraphics[width=7cm]{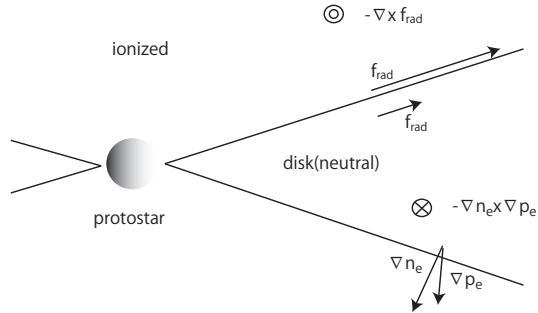}\\
\caption{Schematic view of magnetic field generation in the neighbor of protostar.
}
\label{fig1}
\end{center}
\end{figure}
Now we are able to simulate the generation of magnetic field by
integrating the equation (\ref{eq:B_growth}) by postprocessing
method utilizing the results of radiation hydrodynamics simulation
mentioned above, i.e., the snapshots of the spacial distributions of density,
temperature, velocity and chemical abundances in the accretion envelope.
Then we assess the Biermann battery term by a simple finite difference scheme.
Because of axis symmetry, $\bm{\nabla}n_{\rm e}$ and $\bm{\nabla}p_{\rm e}$ are both
perpendicular to $\bm{e}_\phi$, the base vector of the azimuthal
angle. Thus, the source term $\bm{\nabla}n_{\rm e}\times
\bm{\nabla}p_{\rm e}$ is parallel to $\bm{e}_\phi$(Fig.\ref{fig1}).

We can also calculate the radiation term by ray-tracing from the central
protostar. 
This radiation term $\bm{\nabla} \times \bm{f}_{\rm rad}$ is also directed
to $\bm{e}_\phi$, because of the
axial symmetry again(Fig.\ref{fig1}). We also remark that this term is
expected to be significant on the ionization front, where the shear of $\bm{f}_{\rm rad}$ is large.

Thus, the seed field generated by these source terms is
parallel to $\bm{e}_\phi$. In addition to these source terms, the dynamo term
$\bm{\nabla}\times(\bm{v}\times\bm{B})$ is also parallel to
$\bm{e}_\phi$ as long as $\bm{B}\propto\bm{e}_\phi$ is satisfied. Since we assume
zero field strength at the initial state in the present
simulations, magnetic field is always parallel to $\bm{e}_\phi$
throughout the simulation. It is also worth noting that the dynamo term
reduces to a simple advection term under the assumed symmetry in the
present simulation.

We ignore the contribution from the diffuse radiation emitted
from the surface of the disk, since its contribution to the
photoionization is smaller than the direct radiation from the protostar. 

We use time stepping given by the Courant condition to integrate 
equation (\ref{eq:B_growth}), so that we can handle the advection of
magnetic field. Since the output interval of the radiation hydrodynamics
simulation is larger than the
required time step, we interpolate the data between the output time steps.

\section{Results}
\label{results}
\begin{figure*}
\begin{center}
\includegraphics[width=18cm]{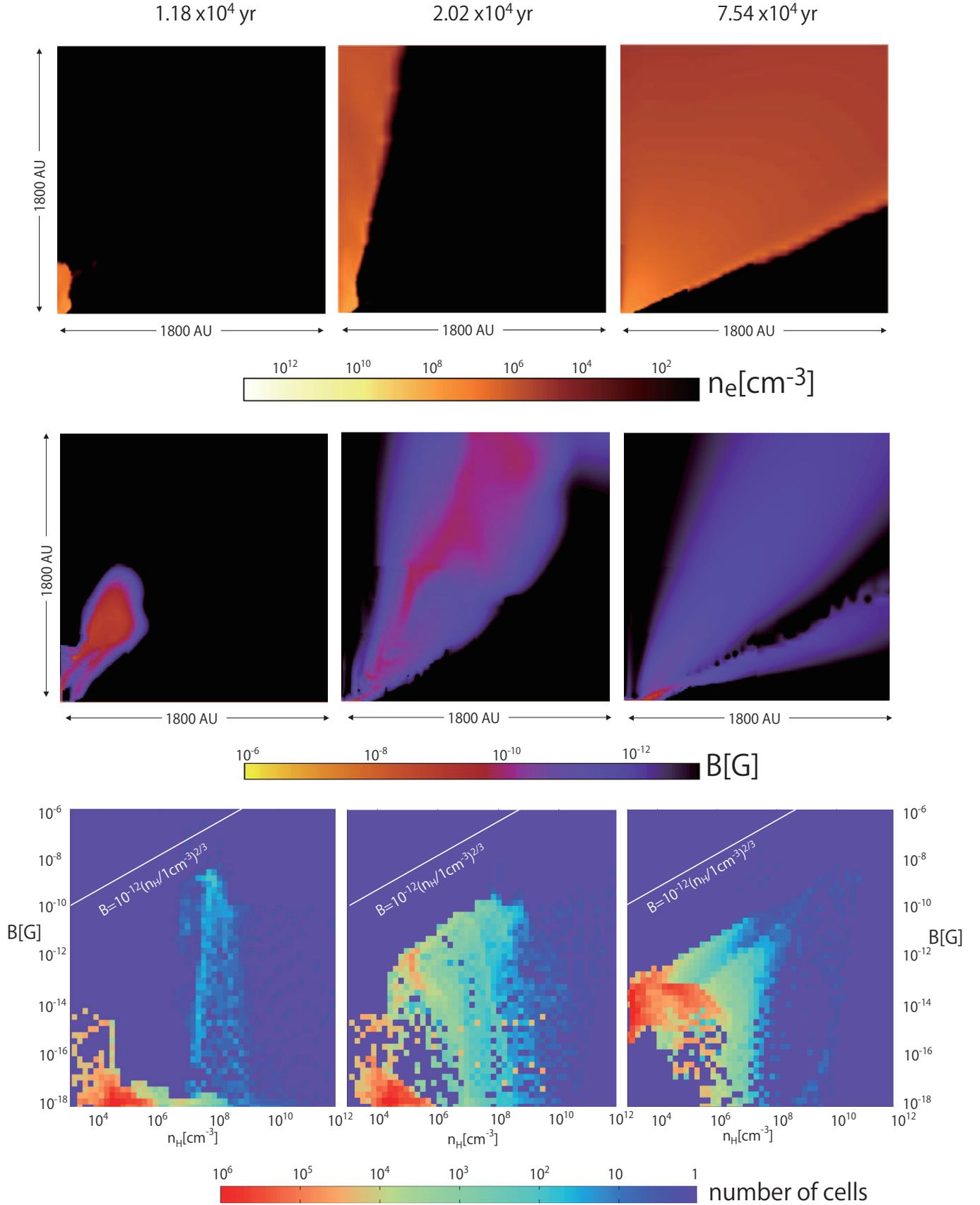}\\
\caption{
Time evolution of magnetic field is presented. Top: electron density
 $n_{\rm e}$ is shown by the color contour. In each panel, protostar is located at the
 left-bottom corner of the box. Three panels corresponds to three
 snapshots. The box is 1800AU on each side.
 Three panels correspond to the snapshots of $1.18\times 10^4$ yrs, $2.02\times 10^4$ yrs, and
$7.54\times 10^4$ yrs after
 the formation of the protostar. middle:magnetic field distribution in the
 neighbor of protostar. Bottom: frequency
 distribution of computational cells on $\log n_{\rm H}{\rm
 [cm^{-3}]}-\log B{\rm [G]}$ plane.
}
\label{fig2}
\end{center}
\end{figure*}

Fig.\ref{fig2} shows the evolution of electron number density and magnetic field around the
protostar. 
Top three panels show the distribution of electron number density on $R-z$ plane of cylindrical coordinate
at $1.18\times 10^4$ yrs, $2.02\times 10^4$ yrs, and
$7.54\times 10^4$ yrs after the formation of the protostar. 
Middle panels show the spatial distribution of magnetic field strength
at the corresponding epochs. The box is 1800AU on a side.
The protostars are located at
the bottom left corners of the boxes, and their masses at the three
epochs are $17.6M_\odot, 23.4M_\odot$ and $39M_\odot$, respectively.
Bottom row shows the volume weighted probability distribution function on $\log n_{\rm H}{\rm
 [cm^{-3}]}-\log B{\rm [G]}$ plane.

The left column corresponds to the time
just after the breakout of ionization front, when the magnetic field is
generated in the very neighbor of the protostar. Roughly speaking, the
ionized/photoheated small bubble (top) is magnetized (middle). 
On the other hand, the field strength is weak in the outer less dense
regions (bottom). 

Another $10^4$yrs later (middle column), an ionization front expand out to this box size to form
an hourglass shaped ionized polar region (top). Magnetic field is generated at
the ionization front, and the generated
magnetic field is transferred to outer less dense regions, riding on the
outflow into the polar direction following $B \propto n_{\rm
H}^{2/3}$ law (bottom), roughly. 

Finally, at $7.54\times 10^4$yrs(right column), the generated field strength is
$\sim 10^{-14}-10^{-13}{\rm G}(n_{\rm H}/10^3{\rm cm^{-3}})^{2/3}$ (bottom). This
magnitude is
still less than the critical field strength shown by white solid line, above which the
fragmentation of the accretion disks around proto-first-stars are suppressed\citep{machida_doi13}.
 
\begin{figure*}
\begin{center}
\includegraphics[width=12cm]{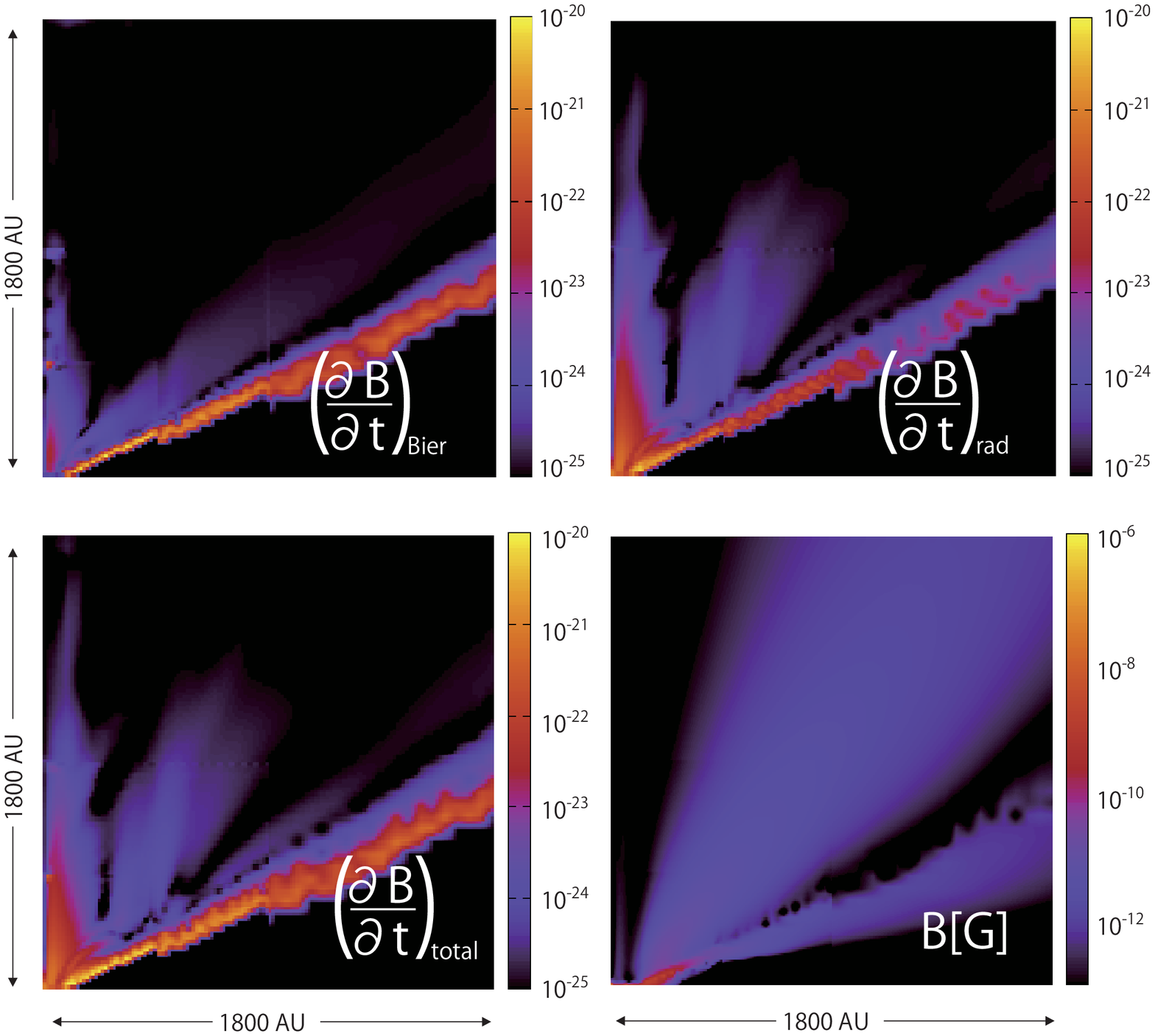}
\caption{Color contour maps of source terms of magnetic field
 generation. The Biermann battery term (top left),
 radiation term(top right), total (bottom left),
 and magnetic field itself (bottom right) at $7.5\times 10^4$yr after the formation of protostar. }\label{fig3}
\end{center}
\end{figure*}

Fig.\ref{fig3} show the Color contour maps of source terms regarding the
magnetic field generation at $7.5\times 10^4$yrs after the formation of
the protostar.  This epoch corresponds to the final snapshot in
Fig.\ref{fig2}.
Top left
panel shows the Biermann battery term, the second term of the left hand side in
equation (\ref{eq:B_growth}), while top right is the radiation term,
i.e. the third term in equation (\ref{eq:B_growth}). The radiation term
is prominent at the ionization front, since the shear field of radiation force is maximized
at the border between the ionized region and the shadowed neutral
region. 
The Biermann battery term is also important at the ionization front,
across which the temperature and the density changes dramatically.
We also remark that the Biermann battery term dominates in more extended regions
(see bottom left, total source term) since the temperature of the
ionized region depends on the distance from the source star very weakly,
while the radiation force $\bm{f}_{\rm rad}$ is inversely proportional
to the square of the distance from the source. Thus, the radiation term
dominates in the neighbor of the protostar.
We also mention that the generated magnetic field has smoother structure
than the source terms, because the gas temperature/density changes with
time and the generated field is transferred to outer less dense regions
(bottom right).

\begin{figure*}
\begin{center}
\includegraphics[width=12cm]{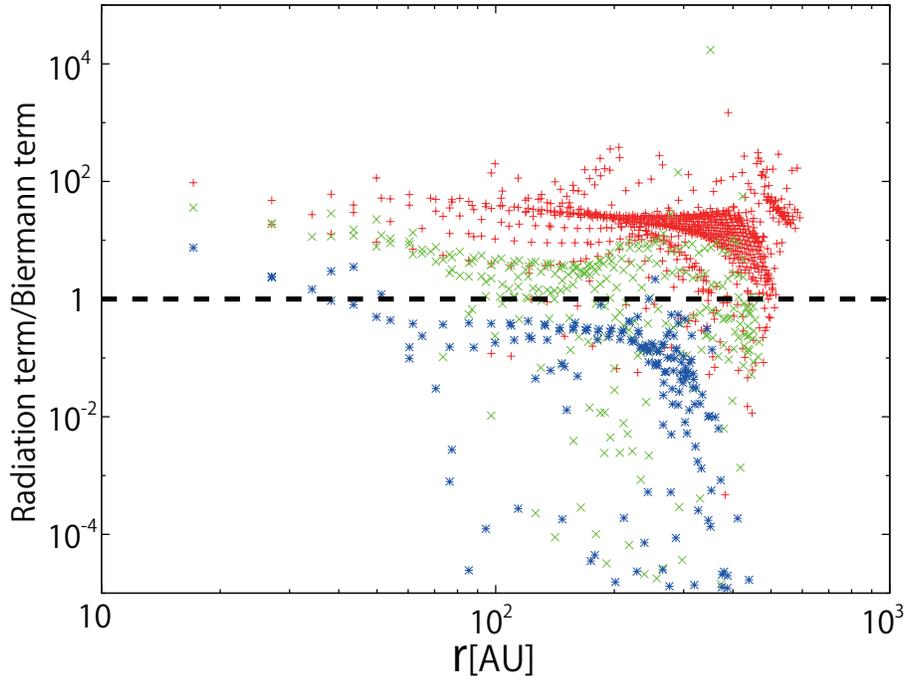}
\caption{The ratio between the radiation source term and the Biermann
 battery term is plotted as a function of distance from the protostar.
Blue stars:$1.18\times 10^4$yr, Green vertices: $2.02\times 10^4$yr and
 Red crosses are for $7.54\times 10^4$yr. These three snapshots correspond the three panels
 in Fig.\ref{fig1}. Points above unity (dashed line) denote the grid cell
 where the radiation term is the dominant source for magnetic field generation.
}\label{fig4}
\end{center}
\end{figure*}

Fig.\ref{fig4} shows the ratio of the radiation term to
the Biermann battery term of each cell in the finest numerical box
(the nested grid scheme are employed in the RHD simulation). Three colors of points corresponds to the
three epochs shown in Fig.\ref{fig2}. It is clear that the Biermann term is more
important than the radiation term just after the break out of the
ionization front(blue stars), while the radiation term dominates in the later
epochs(green and red symbols). This is because the gas is more dynamical in the phase of
I-front break out, which leads to larger gradient of density/pressure,
so is the Biermann battery term. In the later phase, the
density/pressure structure becomes smoother than that in the earlier
phase, and the luminosity of ultraviolet radiation from the protostar
becomes larger. Thus, the relative importance of radiation source term
becomes larger at later epochs.

\section{Discussion \& Conclusion}
\label{discussion}

The magnetic field is generated by two source terms: the radiation term
and the Biermann battery term. The order of magnitude of magnetic field
strength generated via these processes in the neighbor of the central
protostar  can be assessed as follows:
\begin{eqnarray}
B_{\rm rad}&\simeq& \frac{L\sigma_{\nu_{\rm L}}}{8\pi e \Delta r R^2}\Delta t\\
B_{\rm Bier}&\simeq& \frac{c k_{\rm B}T\sin\theta}{e\Delta r^2}\Delta t
\end{eqnarray}
Here, $L$ is the luminosity of the source protostar, $\Delta r$ denotes
the length scale across which $\bm{f}_{\rm rad}$ or the
temperature/density changes, $R$ is the distance from the star, $\Delta
t$ is the duration of field generation, and the $\theta$ represents
the typical angle between $\bm{\nabla}T_{\rm e}$ and $\bm{\nabla}n_{\rm e}$.

The duration $\Delta t$ could be assessed as the time scale that the gas
is sent brazing out distance $R$ by the outflow of typical velocity $v$,
as $\Delta t \equiv R/v$.
Substituting this expression and typical values, we have
\begin{eqnarray}
B_{\rm rad}&\simeq& 10^{-9} {\rm G} \left(\frac{L}{10^{37}{\rm
		     erg/s}}\right)\left(\frac{\Delta r}{10{\rm
		     AU}}\right)^{-1}\left(\frac{R}{100 {\rm
		     AU}}\right)^{-1}\left(\frac{v}{30{\rm km/s}}\right)^{-1}\label{eq:B_o_m}\\
B_{\rm Bier}&\simeq& 10^{-12} {\rm G} \left(\frac{T_{\rm e}}{10^4{\rm K}}\right)\left(\frac{\sin\theta}{0.1}\right)\left(\frac{\Delta r}{10{\rm
		     AU}}\right)^{-2}\left(\frac{R}{100 {\rm
		     AU}}\right)\left(\frac{v}{30{\rm km/s}}\right)^{-1}
\end{eqnarray}
Here we assume $\Delta r = 0.1R$.
These estimated orders of magnitude are consisent with the numerical
results {at the inner high density regions, $n_{\rm H}\ga 10^{8}{\rm cm^{-3}}$.
(see bottom three panels of Fig.\ref{fig2}).}

The generate field strength is $\sim 10^{-13}(n_{\rm
H}/10^3{\rm cm^{-3}}){\rm G}$  at the final stage of the present
simulation, which is account for $\sim 10^{-17}{\rm G}$, at IGM
densities of $z=20$. This magnitude is comparable to the value obtained by
previous estimates at 100pc-1kpc scales \citep{doi}. The coherence
length is also similar with each other, since the outflow will extend
out to the host minihalos of $\sim 100$pc. 
This obtained field strength is less than the critical value above which the fragmentation of
disk is suppressed by two orders of magnitude \citep{machida_doi13}. 
In addition, we also remark that such relatively large magnetic field emerges
after the central protostar grows up to $\ga 20M_\odot$, since the
protostar has to be massive enough to emit ultraviolet radiation. According to a three
dimensional radiation hydrodynamics simulation, the gas disk is heated
by the radiation from the protostar at such epoch, hence the disk is
stabilized against gravitational instability \citep{susa13}. 
Thus, the generated magnetic field in this context seems hardly affect the on-site fragmentation
of the disk directly.

However, as for the second generation star formation, these field
strength could play important roles.
Firstly, recent cosmological magneto-hydrodynamics simulations of first star
formation revealed that the minihalos that host primordial star forming gas clouds are very
turbulent \citep[e.g.][]{turk12}. 
According to the  theoretical model, such turbulent motion at much less
than the Jeans scale could amplify the seed
magnetic field with a dynamical time scale,  and it
inversely cascade into larger scales to affect the dynamics of star
forming cloud at the Jeans scale.
If this mechanism also works in the collapsing gas clouds in the
neighbor of first stars, the seed field formed in the present mechanism
would be important for the formation of second generation stars.

Secondly, we might have underestimated the field strength.
According to the equation (\ref{eq:B_o_m}), we have $B_{\rm rad} \propto
R^{-2}$, assuming $\Delta r \simeq 0.1R$. 
However, the very vicinity around the central protostar is not spatially
resolved in the underlying simulations. With the higher-resolution
simulations resolving the innermost part of the disk, the magnetic
fields generated there should be much stronger than the current
estimates. At the disk surface of $R\sim 10R_\odot$, which is slightly larger
than the radius of a $40M_\odot$ star, the generated magnetic field could
reach $\sim 5\times 10^{-3}$ G
\footnote{It has been pointed out that 
there is an upper limit of the magnetic field strength generated by
radiation drag effects such as the Compton
drag\citep{balbus93,chuzhoy04,silk_langer06}. 
The estimated field
strength here is much larger than the limit. However, there is no
contradiction because the present radaition
effect is not the drag effect.}.
We also can assess the density at the surface of the disk of $R\sim 10R_\odot$ by extrapolating the
results of numerical simulation, where the gas density is approximately proportional to $R^{-1}$. 
Consequently, we obtain $n_{\rm H}\simeq 2\times 10^{11}{\rm cm^{-3}}$. 
The generated magnetic field of $\sim 5\times 10^{-3}$ G at $n_{\rm
H}\simeq 2\times 10^{11}{\rm cm^{-3}}$ will be blown out to outer less dense regions, and result in $B\sim 10^{-8}{\rm G}(n_{\rm H}/10^3{\rm cm^{-3}})^{2/3}$.
This is obviously important for the dynamics of gravitationally
collapsing gas cloud even without the amplification by small scale
dynamo action quoted in the previous paragraph.
However, we remark that higher resolution studies are
necessary to find the actual field strength in the very neighbor of the
proto-first-star, since this is an estimate based upon the extrapolation
of the present results.

{
We also point out the three dimensional effects can also enhance the magnetic
field strength. As shown by recent three dimensional calculations
\citep[e.g.][]{susa13}, the gas disk around the protostar is highly
non-axisymmetric. Such 3D structures induce poloidal component of the magnetic
field, which will result in the dynamo amplification in the disk.
}

\hspace{0.5cm}

In this paper, we assess the magnetic field generated in the
very neighbor of proto-first-star due to the Biermann battery
effect as well as radiation force. As a result, we find that a weak magnetic
field is generated in the inner $\sim 100{\rm AU}-1000{\rm AU}$ region and they are blown
out to the outer less dense regions riding on the
outflows roughly following the $B \propto n_{\rm H}^{2/3}$ low. 
The resultant field strength is $B \sim 10^{-14}-10^{-13}{\rm G}(n_{\rm H}/10^3{\rm
cm^{-3}})^{2/3}$. This field strength can be the seed magnetic field of
the universe and should be important for the next generation star formation, 
 while it hardly affect the dynamics of the on-site first
star formation unless very efficient amplification process is taken into
consideration.

\bigskip
We thank M. Machida, K. Doi and N. Tominaga for fruitful discussions. This
work was supported in part by Ministry of
Education, Science, Sports and Culture, Grant-in-Aid for Scientific
Research (C), 22540295. 


\end{document}